# Hexagonal phase stabilization and magnetic orders of multiferroic Lu$_{1-x}$Sc$_x$FeO$_3$

L. Lin,[1,2] H. M. Zhang,[1] M. F. Liu,[2] Shoudong Shen,[3] S. Zhou,[1] D. Li,[4] X. Wang,[2] Z. B. Yan,[2] Z. D. Zhang,[4] Jun Zhao,[3,5] Shuai Dong,[1 a)] and J. -M. Liu[2 b)]

[1]*Department of Physics, Southeast University, Nanjing 211189, China*

[2]*Laboratory of Solid State Microstructures and Innovation Center of Advanced Microstructures, Nanjing University, Nanjing 210093, China*

[3]*State Key Laboratory of Surface Physics and Department of Physics, Fudan University, Shanghai 200433, China*

[4]*Shenyang National Laboratory for Materials Science, Institute of Metal Research, Chinese Academy of Sciences, Shenyang 110016, China*

[5]*Collaborative Innovation Center of Advanced Microstructures, Fudan University, Shanghai 200433, China*

**Abstract**: Hexagonal LuFeO$_3$ has drawn a lot of research attention due to its contentious room-temperature multiferroicity. Due to the unstability of hexagonal phase in the bulk form, most experimental studies focused on LuFeO$_3$ thin films which can be stabilized by strain using proper substrates. Here we report on the hexagonal phase stabilization, magnetism, and magnetoelectric coupling of bulk LuFeO$_3$ by partial Sc-substitution of Lu. First, our first-principles calculations show that the hexagonal structure can be stabilized by partial Sc substitution, while the multiferroic properties including the noncollinear magnetic order and geometric ferroelectricity remain robustly unaffected. Therefore, Lu$_{1-x}$Sc$_x$FeO$_3$ can act as a platform to check the multiferroicity of LuFeO$_3$ and related materials in the bulk form. Second, the magnetic characterizations on bulk Lu$_{1-x}$Sc$_x$FeO$_3$ demonstrate a magnetic anomaly (probable antiferromagnetic ordering) above room temperature, ~ 425-445 K, followed by magnetic transitions in low temperatures (~167-172 K). In addition, a magnetoelectric response is observed in the low temperature region. Our study provides useful information on the multiferroic physics of hexagonal $R$FeO$_3$ and related systems.

---

a) Email: sdong@seu.edu.cn  b) E-mail: liujm@nju.edu.cn

## I. Introduction

Multiferroics, which exhibit more than one primary ferroic order parameter simultaneously, have been extensively studied in recent years due to the fascinating physics and broad potential applications [1-4]. Unfortunately, so far no many devices associated with the magnetoelectric functions have been reported, concerning the low ferroelectric/magnetic ordering temperatures, small ferroelectric polarizations, and magnetization in most multiferroics. The coexistence of magnetic and electric orders at room temperature is highly required, but in a long period only perovskite $BiFeO_3$ is competent.

The recently discovered hexagonal $LuFeO_3$ ($h$-$LuFeO_3$) may be another room-temperature multiferroic [4-7]. Its crystal structure (see Fig. 1(a)) and multiferroic mechanism are in analogue to hexagonal manganites (e.g. $h$-$YMnO_3$) [8], but quite different from perovskite $BiFeO_3$ or bixbyite-structural $ScFeO_3$ [9,10]. For $h$-$LuFeO_3$, the ferroelectricity appears below the Curie temperature $T_c \sim 1050$ K (for a comparison, $T_c \sim 900$ K for $YMnO_3$) [7,11]. The paraelectric (group $P6_3/mmc$) to ferroelectric (group $P6_3cm$) structural transition can be achieved via the freezing of three phonon modes $\Gamma_2^-$, $K_1$, and $K_3$, as shown in Fig. 1(b) [12,13]. Thus, $h$-$LuFeO_3$ is a kind of improper (geometric) ferroelectric material.

Different from the well-recognized ferroelectricity, the magnetism of $h$-$LuFeO_3$ is under debate. A recent experiment on $h$-$LuFeO_3$ thin films found a considerably high antiferromagnetic (AFM) ordering temperature ~440 K (for comparison, $T_N \sim 80$ K in $YMnO_3$) [14], followed by a second magnetic transition at ~130 K [7]. However, later experiments could not confirm the high temperature antiferromagnetism [15], while only the low temperature transition at ~120-147 K was reported, depending on the substrate and stoichiometry [16]. Nowadays, it is physically interesting and important to solve this conflict and reveal the intrinsic magnetism of $h$-$LuFeO_3$ free of external influences.

Lu-ferrite has two structures: the stablest orthorhombic one ($o$-$LuFeO_3$) and metastable hexagonal one ($h$-$LuFeO_3$) [17,18], as sketched in Fig. 1(a) and 1(c) for comparison. This meta-stability makes experimental study of multiferroicity of bulk $h$-$LuFeO_3$ quite challenging and thus very rare [19-21]. Furthermore, $h$-$LuFeO_3$ has a larger unit cell volume than $o$-$LuFeO_3$, implying that the high pressure synthesis probably not work for $h$-$LuFeO_3$. Although $h$-$LuFeO_3$ films can be stabilized on some substrates, its intrinsic physical properties may be seriously affected due to the strain effects, interface/surface effects, and widely existing defects in films [16]. Thus it is crucial to perform experimental studies using high quality bulk samples. However, different from $R$MnO$_3$ in which the hexagonal phase becomes stable over the orthorhombic one when $R^{3+}$ is small, the orthorhombic phase is always stable in $R$FeO$_3$, considering Lu$^{3+}$ is already the smallest rare earth ion.

Recently, a stable hexagonal structure was reported in scandium (Sc) substituted LuFeO$_3$ [22,23]. There are three interesting questions arisen from this work. First, why the substitution of Sc can stabilize the hexagonal structure, noting that neither ScFeO$_3$ nor LuFeO$_3$ prefers the *P*6$_3$*cm* hexagonal one. Second, how those physical properties, especially the multiferroicity, are unaffected or affected upon such a substitution. Third, is there any high temperature magnetic transition upon the substitution? In Ref. [22], no magnetic measurement above room temperature was performed, although a weak ferromagnetic order below $T_R \sim 162$ K was detected. In Ref. [23], the low temperature magnetic transition was investigated using neutron scattering, and a paramagnetic behavior between 200 K to 400 K was claimed, without magnetic data above 400 K.

In this work, we have performed both theoretical and experimental studies on the magnetic, ferroelectric, and dielectric properties of bulk Lu$_{1-x}$Sc$_x$FeO$_3$. First, our density functional theory (DFT) calculations explained the stability of hexagonal structure upon appropriate Sc substitution. Second, our calculations also found that both the magnetism and ferroelectricity are almost unaffected by the substitution. Third, our experiments found a magnetic anomaly (maybe an AFM ordering) at $T_A \sim 425$-445 K, followed by the low temperature magnetic transitions at $T_N \sim 167$-172 K. The last but not the least, our pyroelectric polarization measurements found a magnetoelectric response in the low temperature region, which was predicted for hexagonal manganites but has never been directly observed [24].

## II. Density functional theory calculations

II.A. Method

Our first-principles calculations were performed using the projected augmented wave (PAW) pseudopotentials implemented in the Vienna *ab-initio* Simulation Package (VASP) [25,26]. The electron interactions were described using the Perdew-Burke-Ernzerhof function revised for solids (PBEsol), and the generalized gradient approximation plus *U* (GGA+*U*) method was adopted to treat the exchange and correlation of electrons. The choice of PBEsol can give a more accurate description of structures than traditional PBE. The original PBE is biased towards a correct description of molecules [27], while the PBEsol generalized gradient approximation can improve equilibrium properties of densely packed solids and their surfaces [28]. The Dudarev implementation was adopted to treat the on-site Coulomb $U_{eff}=U-J=4.0$ eV applied to the 3*d* electrons of Fe [29]. The plane-wave energy cutoff was set to be 500 eV and the Monkhorst-Pack *k*-point mesh is 5×5×3 for the hexagonal structures and 7×7×5 for the orthorhombic structure.

Regarding the magnetism, various collinear spin orders, including ferromagnetic (FM), A-type AFM (A-AFM), G-type AFM (G-AFM, for orthorhombic structure only), and the 120° noncollinear spin order (named as Y-AFM here, for hexagonal structure only), were calculated. Spin-orbit coupling was not taken into account considering its weak effect to phase stability. Thus the absolute axes of spins are not involved in our calculations. Lattice relaxations, including the lattice constants and inner atomic positions, were performed with magnetism until the Hellman-Feynman forces less than 0.01 eV/Å. After the relaxations, energies were calculated and the standard Berry phase method was employed to calculate the ferroelectric polarization [30].

II.B. Results

Before the study on substituted systems, the pure Lu-ferrites are checked. According to previous works [31], the magnetic ground state of *o*-LuFeO$_3$ is G-AFM, which is also confirmed in our DFT calculation. The energy of this state will be used as a reference for comparison.

For *h*-LuFeO$_3$, the energies of various magnetic states were calculated, as listed in Table I. Among various magnetic states, the non-collinear Y-AFM is the lowest in energy, consistent with experimental results [7, 32]. Even though, this energy is still higher than that of the G-AFM state in the orthorhombic structure, implying that *o*-LuFeO$_3$ is stabler than *h*-LuFeO$_3$. These results agree with experimental facts. Furthermore, the relaxed lattice constants also agree with the experimental values quite well [5]. All these consistencies between our calculations and experiments guarantee the reliability of the following calculations on substituted systems.

Then a half-substituted system (Lu$_{0.5}$Sc$_{0.5}$FeO$_3$) is taken into account and both the orthorhombic and hexagonal structures are considered. In minimal unit cells for *o*-LuFeO$_3$ (four Fe's) and *h*-LuFeO$_3$ (six Fe's), there are three and five independent configurations of substitution respectively, as sketched in Fig. 2(a-b). The calculated energies for Sc-substituted *h*-LuFeO$_3$ and *o*-LuFeO$_3$ are displayed in Fig. 2(c). Strikingly, all the five configurations of half Sc-substituted hexagonal systems with the Y-AFM order (and only this magnetic order) are lower in energy than the three configurations of orthorhombic one with G-AFM order. In particular, the energy difference between these two structures reaches 2.4-31.4 meV/Fe depending on the detailed configurations, as shown in Fig. 2(c). As a conclusion of the energy comparison, $Lu_{0.5}Sc_{0.5}FeO_3$ prefers to crystallize in the hexagonal structure instead of the orthorhombic one, offering a chance to further study the hexagonal $R$FeO$_3$ in the bulk form.

Then, does the hexagonal Sc-substituted LuFeO$_3$ show similar or discrepant magnetoelectric properties in comparison with original $h$-LuFeO$_3$? First, as shown in Fig. 2(c), the non-collinear Y-AFM, is stable in both the original $h$-LuFeO$_3$ and the substituted one. Then the nearest-neighbor (NN) exchange interactions in $h$-LuFeO$_3$ and $h$-Lu$_{0.5}$Sc$_{0.5}$FeO$_3$ were calculated for comparison, which are the key coefficients to determine the magnetic transition temperatures. For simplicity, only the NN in-plane and interlayer exchange interactions are considered, as indicated in Fig. 2(d). The energy for exchange interactions between Fe sites can be written as:

$$H_{ex} = J_1 \sum_{<ij>} S_i \cdot S_j + J_2 \sum_{<ik>} S_i \cdot S_k, \qquad (1)$$

where $J_1$ is the in-plane superexchange between two NN Fe spins, $J_2$ is the interlayer superexchange between two NN Fe spins. Here $|S|$=5/2 is adopted considering the high-spin fact of $3d^5$ electron configuration. In fact, in the $P6_3cm$ structure, due to the in-plane trimerization of Fe triangles, there are two types of in-plane NN Fe-Fe pairs (shorter $vs$ longer). And due to the tilting of oxygen bipyramids, there are two types of inter-layer exchanges. Here, for simplicity, these fine differences are neglected and only the average in-plane and inter-layer exchanges are considered.

Using the optimized structures for Y-AFM, the exchanges $J_1$ and $J_2$ can be extracted by comparing the energies of various states (FM, A-AFM, and Y-AFM), as shown in Fig. 2(d). As expected for a quasi-layered structure, the in-plane interaction $J_1$ is much stronger than $J_2$. Our calculation shows that $J_2$ keeps almost unchanged upon the Sc substitution regardless of the configuration details. In contrast, there is a slight enhancement of $J_1$ from 6.22 meV (in $h$-LuFeO$_3$) to 6.94-7.25 meV (in $h$-Lu$_{0.5}$Sc$_{0.5}$FeO$_3$), implying that the corresponding magnetic transition temperature(s) will be probably increased a little upon the Sc substitution. Since $Sc^{3+}$ is smaller than $Lu^{3+}$, the substitution will shrink the lattice constants, especially the in-plane ones (e.g. 1.7%-2.0% smaller for the lattice constants $a$-$b$ and only 0.4%-1.1% smaller for the lattice constant $c$ in the DFT calculation). Thus the shortened Fe-O-Fe bond lengths enhance the in-plane superexchanges. The shrunk lattice is advantageous to stablize the hexagonal phase according to the experience of $R$MnO$_3$.

Second, the ferroelectricity has also been investigated. The optimized crystal structures of paraelectric and ferroelectric phases for $h$-Lu$_{0.5}$Sc$_{0.5}$FeO$_3$ are plotted in Fig. 3(a-b). Very similar to the distortions in $h$-LuFeO$_3$, the tilting of FeO$_5$ bipyramids and the up-down-down displacement of $Lu^{3+}$ ($Sc^{3+}$) are prominent, which contribute to the polarization. The calculated ferroelectric polarization by the Berry phase method is summarized in Fig. 3(c). The values for all the five configurations are close to the one of pure $h$-LuFeO$_3$, implying that the saturated ferroelectric polarization of $h$-Lu$_{0.5}$Sc$_{0.5}$FeO$_3$ should be very similar to that of $h$-LuFeO$_3$. In addition, the energy barrier between the paraelectric state (with Fe-trimerization) and ferroelectric state, which is an indication of ferroelectric transition temperature, has also been calculated, as shown in Fig. 3(d), whose heights are slightly lower in $h$-Lu$_{0.5}$Sc$_{0.5}$FeO$_3$. Thus, the ferroelectric transition temperature should be decreased upon the Sc substitution. Considering the original very high ferroelectric transition temperature of $h$-LuFeO$_3$ (1050 K), a moderate transition transition temperature may be advantaged for room temperature ferroelectric operations.

The electronic structures of $h$-LuFeO$_3$ and Sc-substituted LuFeO$_3$ have also been calculated, as presented in Fig. 3(e-f). The band gap $E_g$ is ~1.3-1.4 eV for both $h$-LuFeO$_3$ and $h$-Lu$_{0.5}$Sc$_{0.5}$FeO$_3$, the latter of which is slightly lower (~0.1 eV). The value for $h$-LuFeO$_3$ is consistent with previous prediction [33], which is actually under-estimated compared with experimental one due to the well-known drawback of DFT. In other words, our DFT calculation suggests that electronic structure of $h$-Lu$_{0.5}$Sc$_{0.5}$FeO$_3$ should be very similar to that of $h$-LuFeO$_3$.

To summarize our DFT calculations, the half Sc-substitution can stabilize the hexagonal phase of LuFeO$_3$ and keeps the multiferroic properties almost unaffected, which provides an idea model system to experimentally explore the magnetoelectricity of hexagonal $R$FeO$_3$ in the bulk form.

## III. Experiments

Meanwhile, we have synthesized the whole Lu$_{1-x}$Sc$_x$FeO$_3$ series and performed systematic characterizations on these samples.

III.A. Method

Polycrystalline Lu$_{1-x}$Sc$_x$FeO$_3$ samples were prepared by standard solid state reaction method [22]. In detail, the stoichiometric and high-purity starting material Lu$_2$O$_3$, Fe$_2$O$_3$, and Sc$_2$O$_3$ were first thoroughly ground, and sintered at 1000 °C for 12 hours. Then the powder was pelletized and sintered at 1200 °C for 24 hours with intermittent grinding step. To investigate the chemical binding and valence state, X-ray photoelectron spectrometer (XPS) (PHI 5000 VersaProbe by UlVAC-PHI, Inc.) and Mössbauer spectrum were measured. The electron dispersion spectrum (EDS) (S-3400N II by Hitachi, Inc.) was used to analyze the chemical composition, and thermogravimetry (TG) measurement was also carried out to investigate the possible oxygen vacancy in samples.

To measure dielectric and ferroelectric properties, each sample was first polished into a disk of diameter of ~3.0 mm, thickness of ~0.3 mm, and then coated with Au on two sides as electrodes. The temperature ($T$)-dependence of dielectric constant ($\varepsilon$) was measured using the HP4294A impedance analyzer integrated with Physical Property Measurement System (PPMS) (Quantum Design, Inc.). The measurement of specific heat ($C_p$) was carried out using the PPMS in the standard procedure. The ultraviolet absorption spectrum was carried out in the phonon energy range of 1.55-5.0 eV (240-800 nm) with a spectra resolution 1.0 nm by UV-Vis Spectrometer (UV-2450 by SHIMADZU CO. INC).

For magnetic property below 300 K, the magnetization ($M$) as a function of temperature under zero-field cooling (ZFC) and field cooling (FC) sequences was measured by the Superconducting Quantum Interference Device Magnetometer (SQUID) (Quantum Design, Inc.), with a measuring/FC magnetic field of 1 kOe. In the high temperature range (300-600 K), the ZFC-FC $\chi$-$T$ curves were measured in a high vacuum state using a vibrating sample magnetometer (VSM) with the oven option. To probe the ferroelectric polarization ($P$) at low temperature region, the pyroelectric current method was used, and extrinsic contributions, such as the de-trapped charges, were carefully reduced to a minimal level [34-35]. The temperature-dependent polarization under magnetic field ($H$) was collected at a warming rate of 4 K/min with a fixed magnetic field during the whole process. The poling electric field ($E$) before the pyroelectric measurement is 6.7 kV/cm.

III.B. Results

The X-ray diffraction (XRD) patterns for $Lu_{1-x}Sc_xFeO_3$ samples are shown in Fig. 4(a). The single phase $o$-LuFeO$_3$ and bixbyite ScFeO$_3$ were obtained at $x$=0 and 1, respectively, while all the reflections for each sample can be properly indexed by the $o$-LuFeO$_3$, $h$-LuFeO$_3$, bixbyite ScFeO$_3$, or a combination of them. With increasing $x$, the intensities of Brag peaks for orthorhombic phase gradually drop down and those for hexagonal phase emerge since $x$=0.2. The pure hexagonal phase was obtained at $x$=0.5 and 0.6 within the XRD resolution limit. The bixbyite phase gradually dominates upon further increasing $x$ (e.g. to 0.8 and over). These results confirm that appropriate Sc substitution around $x$=0.5 can indeed stabilize the hexagonal phase [22]. For clarity, the slowly-scanned XRD patterns for samples $x$=0.5 and 0.6 are plotted in Fig. 4(b), and the standard index for $h$-LuFeO$_3$ and bixbyite ScFeO$_3$ are inserted for reference. It can be clearly seen that the $x$=0.5 sample is in perfect hexagonal phase (no detectable impurity within the resolution of instrument), while very tiny amount of ScFeO$_3$ impurity exists in the $x = 0.6$ sample as magnified in the inset of Fig.4(b). We also carried out the refinement for the sample $x$=0.5 using the Fullprof program [36]. The refined result is in good agreement with the hexagonal structure, giving the lattice constants $a$=5.8664 Å and $c$=11.7141 Å with reliability parameters $R_{wp}$=7.79%, $R_p$=5.62%, and $R_F$=5.20%.

To further examine the composition of sample $x$=0.5, e.g. cation/anion non-stoichiometry, a series of measurements were performed to investigate the as-obtained sample. First, the EDS measurement shows that the atom ratio of elements Lu: Sc: Fe: O is 10.81: 10.00: 19.85: 59.34 (Fig.5(a)), which is very close to the nominal composition considering the instrumental resolution (measurement error ±1%). Furthermore, the Mössbauer spectrum shows an excellent single doublet without any other obvious sets of peaks, as shown in Fig.5(b), indicating sole type of Fe$^{3+}$ state. The XPS core level spectra of Fe 2$p$ is presented in Fig.5(c). The measured Fe 2$p_{3/2}$ peak is located at 710.80 eV, consistent with reported values for Fe$^{3+}$ [37]. In addition, the satellite observed at 718.82 eV is a diagnostic signature of Fe$^{3+}$ [38]. In other word, both the Mössbauer spectrum and XPS results suggest that the Fe ion exists as Fe$^{3+}$ in the $x$=0.5 sample. Similar results are also obtained for the $x$=0.6 sample, as shown in Fig. 5(c). The TG measurement shows that both samples have weight loss in the temperature range 300-330 K, which can be assigned to the physical adsorption of water molecules on the samples [39]. After releasing the physical adsorption, the oxygen vacancies can be refilled with oxygen, causing a weight gain during 330-370 K [39]. With further increasing temperature, there is smooth and slight increasing, without any anomaly till 1010 K which corresponds to the ferroelectric transition. This transition is slightly lower than the one of LuFeO$_3$ thin film [7], consistent with our DFT calculation. The oxygen vacancy concentration can be roughly deduced from the refilled oxygen amount, which is ~2.8% atomic ratio for the $x$=0.5 sample and ~3.3% atomic ratio for the $x$=0.6 sample.

Subsequently, the temperature dependent magnetism was measured. Fig. 6(a) shows the ZFC and FC magnetic susceptibility ($\chi$) for the $x$=0.5 sample. With decreasing temperature, the curves of ZFC and FC modes start to separate from a magnetic transition temperature $T_{N1}$ ~167 K, followed by a weak anomaly at $T_{f1}$~ 162 K, and then converges below ~10 K. Similar magnetic characters are also apparent in the $x$=0.6 sample (Fig. 6(b)) at a higher $T_{f2}$=168 K and $T_{N2}$ ~172 K. The temperature $T_N$ is the highest in hexagonal $R$MnO$_3$ and $R$FeO$_3$ families [40-41]. This enhancement of $T_N$ is an expected consequence of the enhanced $J_1$ as predicted in above first-principles calculation. In addition, the splitting of the ZFC and FC curves usually suggests the existence of some kind of "magnetic nanoparticle" system at low temperatures: either a magnetic phase separated state, cluster-glass, or spin-glass, etc [42]. In this case, the splitting is attributed to the unusual spin-reorientation as a consequence of coexisting of two magnetic orientations developed within the *ab* plane [23].

The most interesting conflict in *h*-LuFeO$_3$ is the magnetism above room temperature [7, 15]. Here, our VSM measurement on bulk samples shows that there is indeed a magnetic anomaly above room temperature, evidenced by the temperature dependent $\chi$, as shown in Fig. 6(c-d). For both samples, the Curie-Weiss behaviors are observed above 450 K, with Curie-Weiss temperature ($\theta_{cw}$) -1070 K for *x*=0.5 and -980 K for *x*=0.6 extracted from the linear fitting of inverse magnetic susceptibility, as shown in Fig.6(e), indicating a probable AFM phase transition. The anomaly occurs at $\sim T_{A1}$=445 K, and $\sim T_{A2}$=425 K as identified from the 1/$\chi$ fitting (Fig.6(e)) as well as the $d\chi/dT$ curves (Fig.6(f)).

It is crucial to check the reliability as well as the physical origin of this magnetic anomaly. The $\chi$-*T* curve for the *x*=0.5 sample suggests a weak FM signal below $T_{A1}$, which is possible the spin-canting moment from AFM background as proposed by Wang et al. [7]. Once the AFM order is established, the weak FM canting can be driven by the Dzyaloshinskii-Moriya interaction [32,43]. Of course, the oxygen vacancies may also contribute to the magnetic signal in this temperature region [44]. As shown in Fig. 6(g), the magnetic hysteresis loops at 300 K indeed show a nonzero residue *M* and nonzero coercivity for the *x*=0.5 sample but faint signal for the *x*=0.6 sample. This *M-H* loops seems to conflict with the oxygen vacancies driven magnetism, since the oxygen vacancies are higher in the *x*=0.6 sample. With decreasing temperature, the hysteresis loop becomes more clear below $T_N$ but the coercivity almost disappears below 10 K (the convergence point of FC/ZFC curves), as shown in Fig.6(h).

Besides oxygen vacancies, it is also necessary to check possible contribution from other impurities if any. A careful analysis can exclude their contribution to this transition at $T_A$. First, the magnetic transition at $T_A$ can not be due to the *o*-LuFeO$_3$ which has a much higher AFM transition ~628 K [17]. Second, other possible impurities, e.g. LuFe$_2$O$_4$, Fe$_3$O$_4$, or bixbyite ScFeO$_3$, if available (non-detectable in our XRD data), will not contribute to this transition at $T_A$ either considering their magnetic transitions (~240 K for LuFe$_2$O$_4$ [45], ~848 K for Fe$_3$O$_4$ [21], and ~545 K for bixbyite ScFeO$_3$ [9]). Third, our XRD refinement and Mössbauer spectrum show a pure hexagonal phase.

The UV absorption spectrum was measured for the *x*=0.5 sample at 300 K. As shown in Fig. 7(a), two broad peaks show up at ~2.5 eV and ~4.8 eV, consistent with the optical property of *h*-LuFeO$_3$ film [33]. An optical band gap of 1.75 eV is extracted (Fig. 6(b)), which is slightly lower than the value (2.0 eV) of *h*-LuFeO$_3$ [46], as expected by above DFT calculations. However, the peak at 2.9 eV appeared in films [33] does not appear in our sample. Similar band gap is also found in the $x=0.6$ sample.

Then our concern goes to the ferroelectricity, dielectric property, and magnetoelectric coupling in the Sc-substituted LuFeO$_3$ samples. As stated before, the samples are already ferroelectric at room $T$, which is driven by the freezing of the three collective phonon modes (Fig. 1). In this sense, $h$-Lu$_{1-x}$Sc$_x$FeO$_3$ ($x\sim$0.5-0.6) is already a room temperature ferroelectric material. Since the pure ferroelectricity at room temperature has been studied [7,23], here our main interest is the magnetoelectric coupling. The high-precision pyroelectric current method, which is reliable in the low temperature region, was used to measure the change of polarization (not the total polarization) below $T_N$. Taking the $x=0.5$ sample for example, the evaluated polarization $\Delta P(T)$ is plotted in Fig. 8(a), and its ensuing temperature $T_c$ coincides with $T_{f1}$. The dielectric constant also shows a weak anomaly around $T_{f1}$ [22]. Considering the coincidence among dielectric, pyroelectric, magnetism, and specific heat (Fig. 8(b)), it is reasonable to conclude the spin-reorientation at $T_{f1}$ leads to an additional ferroelectric transition, which should be magnetoelectrically active. The saturated $\Delta P$ reaches $\sim$135 $\mu$C/m$^2$ below 80 K. Considering the polycrystalline fact, the intrinsic value should be larger for one order of magnitude. Even though, it should be noted that the obtained polarization here (in fact the change of polarization) should be much smaller than the calculated one which is the saturated full polarization.

Besides, the polarization has a weak response to magnetic stimulus below $T_{f1}$, as shown in Fig. 8(c). In fact, the giant structural distortion, i.e. significant magneto-elastic coupling, was once reported in multiferroic hexagonal (Y$_{1-x}$Lu$_x$)MnO$_3$ accompanying the AFM transition at 75 K despite its very high ferroelectric transition temperature ($\sim$900 K) [24]. Considering the physical similarity between hexagonal $R$MnO$_3$ and $R$FeO$_3$, the magnetoelectric signals at $T_{f1}$ observed in our experiment should also be due to the spin-lattice coupling. To our best knowledge, the direct pyroelectric polarization signal around magnetic transitions in hexagonal $R$MnO$_3$ and $R$FeO$_3$ has not been studied before, which is a fingerprint of magnatoelectricity.

## V. Conclusion

In summary, we have concluded that bulk hexagonal $R$FeO$_3$ can be stabilized by appropriate Sc substitution, confirmed by both the first-principles calculation and experiment. A magnetic anomaly occurring above 425 K (a probable antiferromagnetic transition), has been detected in the Sc-substituted samples, followed by low magnetic transition transitions around 167~172 K. The origin of high temperature magnetism is argued to be intrinsic by excluding possible extrinsic contributions, while certainly more direct evidences (e.g. neutron studies) are needed to resolve the magnetic state in this high temperature region. An unambiguous coupling between ferroelectric order and magnetic order is observed accompanying the spin reorientation at low temperature. These observations open a new possibility to hunting for room temperature multiferroics in other hexagonal $R$FeO$_3$ system and related materials.

**Acknowledgment**

Work was supported by the 973 Projects of China (Grant No. 2015CB654602), the National Natural Science Foundation of China (Grant Nos. 11234005, 11374147, 11504048, 51332006, 51322206), the Priority Academic Program Development of Jiangsu Higher Education Institutions, China Postdoctoral Science Foundation funded project (Grant No. 2015M571630), and Post-doctoral Science Fund of Jiangsu Province (Grant No. 1402043B). S.S and J.Z. Were supported by the National Natural Science Foundation of China (Grant No. 91421106)

**TABLE I.** The DFT energies (in unit of meV/Fe) of various magnetic states for orthorhombic and hexagonal LuFeO$_3$. The G-AFM state of orthorhombic structure is taken as the reference.

|  | G-AFM | FM | A-AFM | Y-AFM |
|---|---|---|---|---|
| $o$-LuFeO$_3$ | 0.0 | 214.9 | 143.6 | - |
| $h$-LuFeO$_3$ | - | 223.8 | 69.7 | 50.1 |

**Captions of figures**

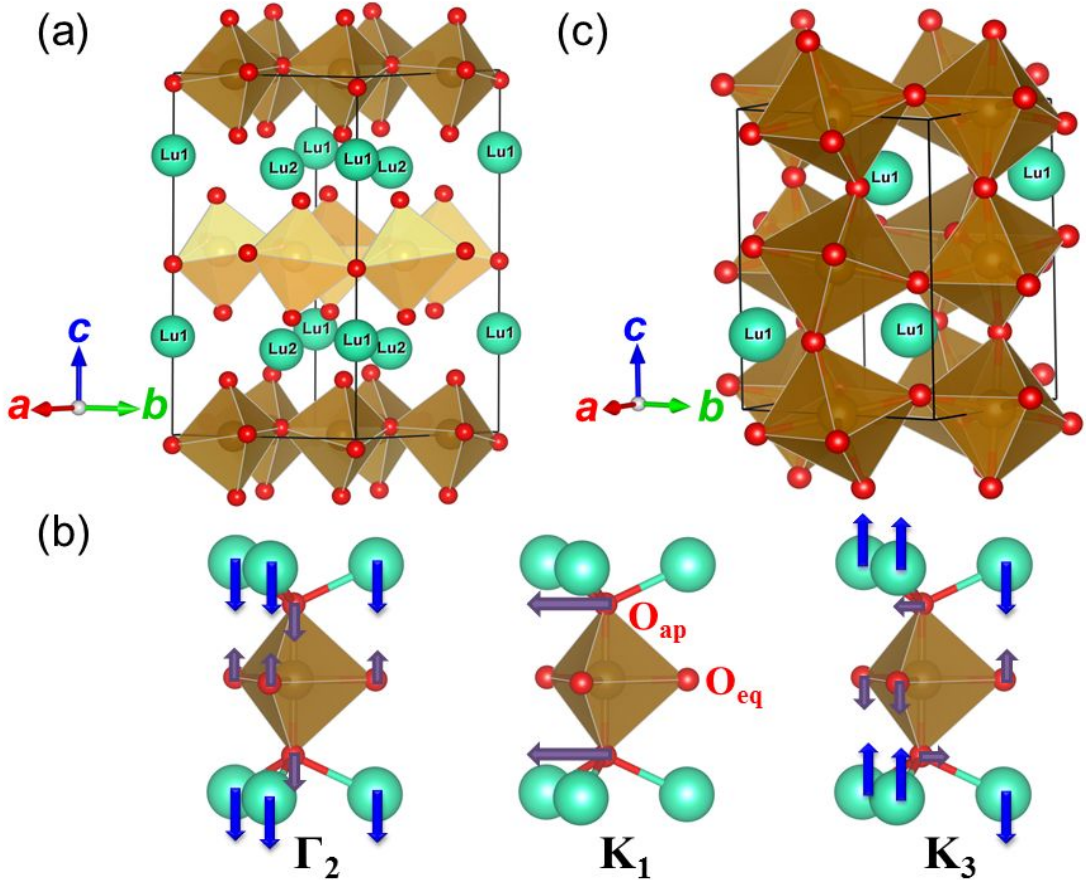

**Fig. 1.** (Color online) (a) Crystal structure of hexagonal LuFeO$_3$. In $h$-LuFeO$_3$, each Fe$^{3+}$ is caged by five O$^{2-}$, forming a FeO$_5$ trigonal bipyramid (TBP). (b) Displacement patterns in the three phonon modes ($\Gamma_2^-$, $K_1$, and $K_3$) related to the $P6_3/mmc$ to $P6_3cm$ structural transition. (c) Crystal structure of orthorhombic LuFeO$_3$. In $o$-LuFeO$_3$, Fe$^{3+}$ ions occupy the octahedral (OCT) sites.

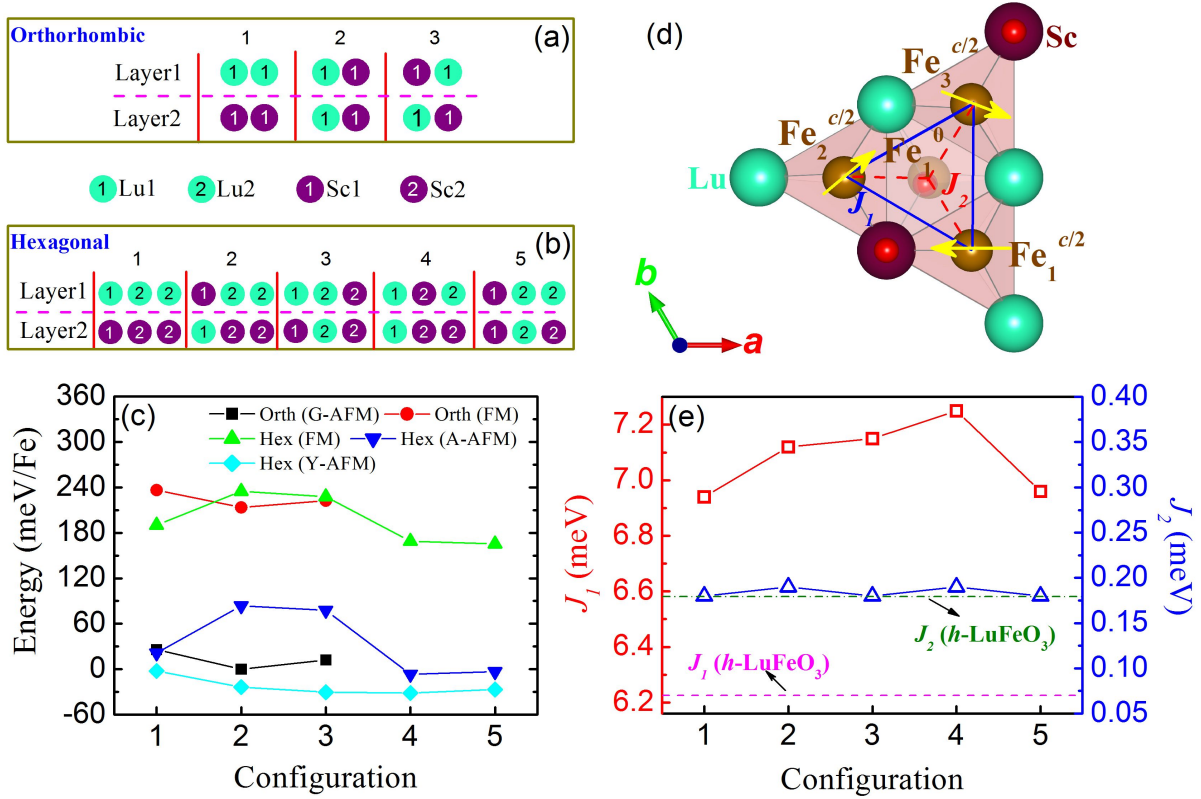

**Fig. 2.** (Color online) The schematic diagram of $Lu^{3+}$ ($Sc^{3+}$) configuration for (a) $o$-$Lu_{0.5}Sc_{0.5}FeO_3$ and (b) $h$-$Lu_{0.5}Sc_{0.5}FeO_3$. The "configuration" concept is explained in the main text. (c) The $Lu^{3+}$ ($Sc^{3+}$) configuration dependence of energies for hexagonal $Lu_{0.5}Sc_{0.5}FeO_3$ (with various magnetic orders: FM, A-AFM, and noncollinear Y-AFM), and for orthorhombic $Lu_{0.5}Sc_{0.5}FeO_3$ (with various magnetic orders: FM, G-AFM), respectively. (d) Illustration of NN superexchange paths $J_2$ between $Fe_1^0$ at $z/c=0$ and three neighboring $Fe_1$, $Fe_2$, and $Fe_3$ at $z/c=1/2$. The superexchange paths $J_1$ between NN coplanar $Fe_1$, $Fe_2$, and $Fe_3$ are also shown. (e) The extracted exchanges $J_1$ and $J_2$ for $h$-$Lu_{0.5}Sc_{0.5}FeO_3$. The corresponding $J_1$ and $J_2$ for $h$-$LuFeO_3$ are also indicated in lines for comparison.

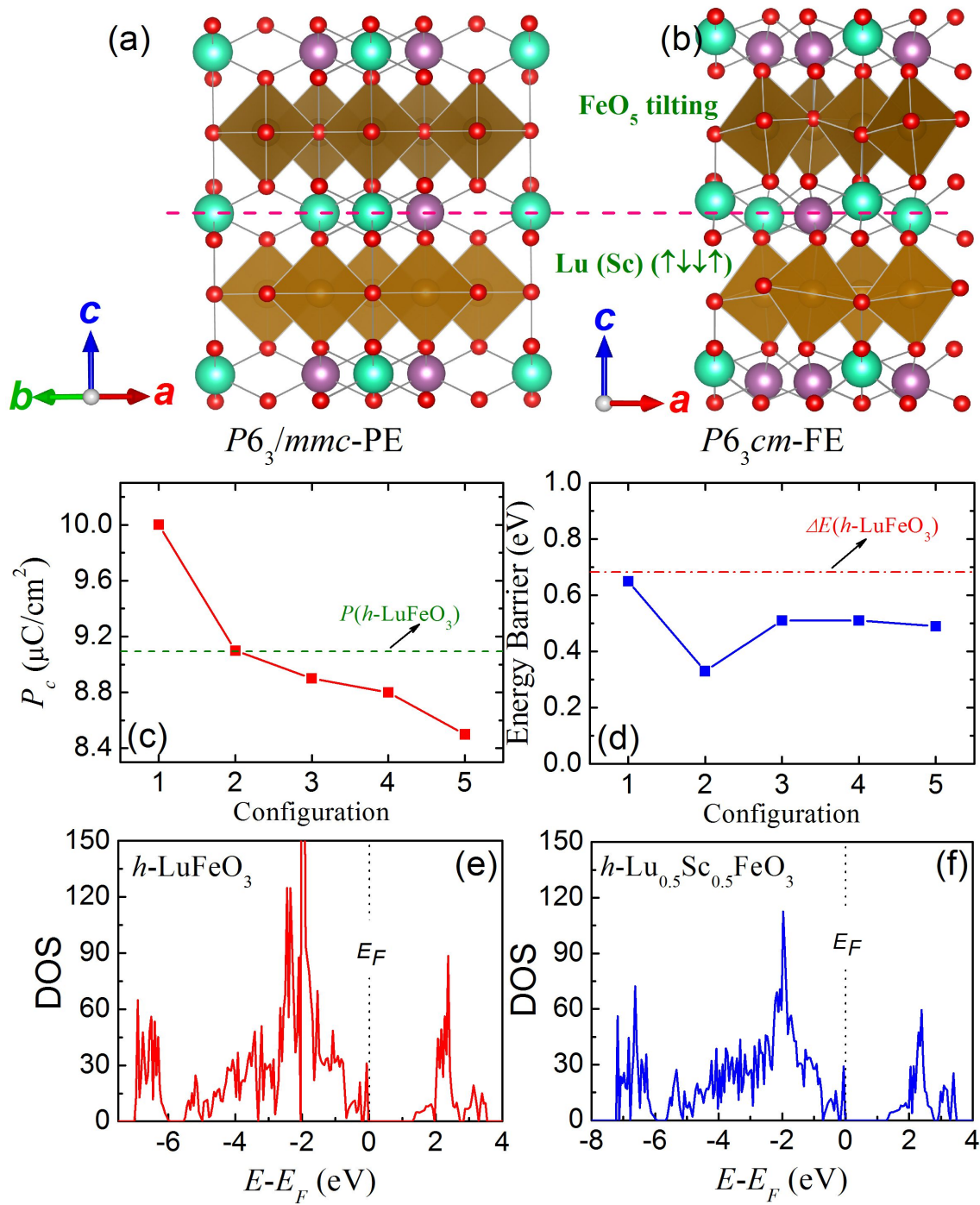

**Fig. 3.** (Color online) The optimized crystal structure (configuration 4) of (a) paraelectric and (b) ferroelectric states of $h$-$Lu_{0.5}Sc_{0.5}FeO_3$. (c) The calculated polarization by Berry Phase for $h$-$Lu_{0.5}Sc_{0.5}FeO_3$. (d) The calculated energy barriers for ferroelectric switch for $h$-$Lu_{0.5}Sc_{0.5}FeO_3$. The density of states of (e) $h$-$LuFeO_3$ and (f) $h$-$Lu_{0.5}Sc_{0.5}FeO_3$.

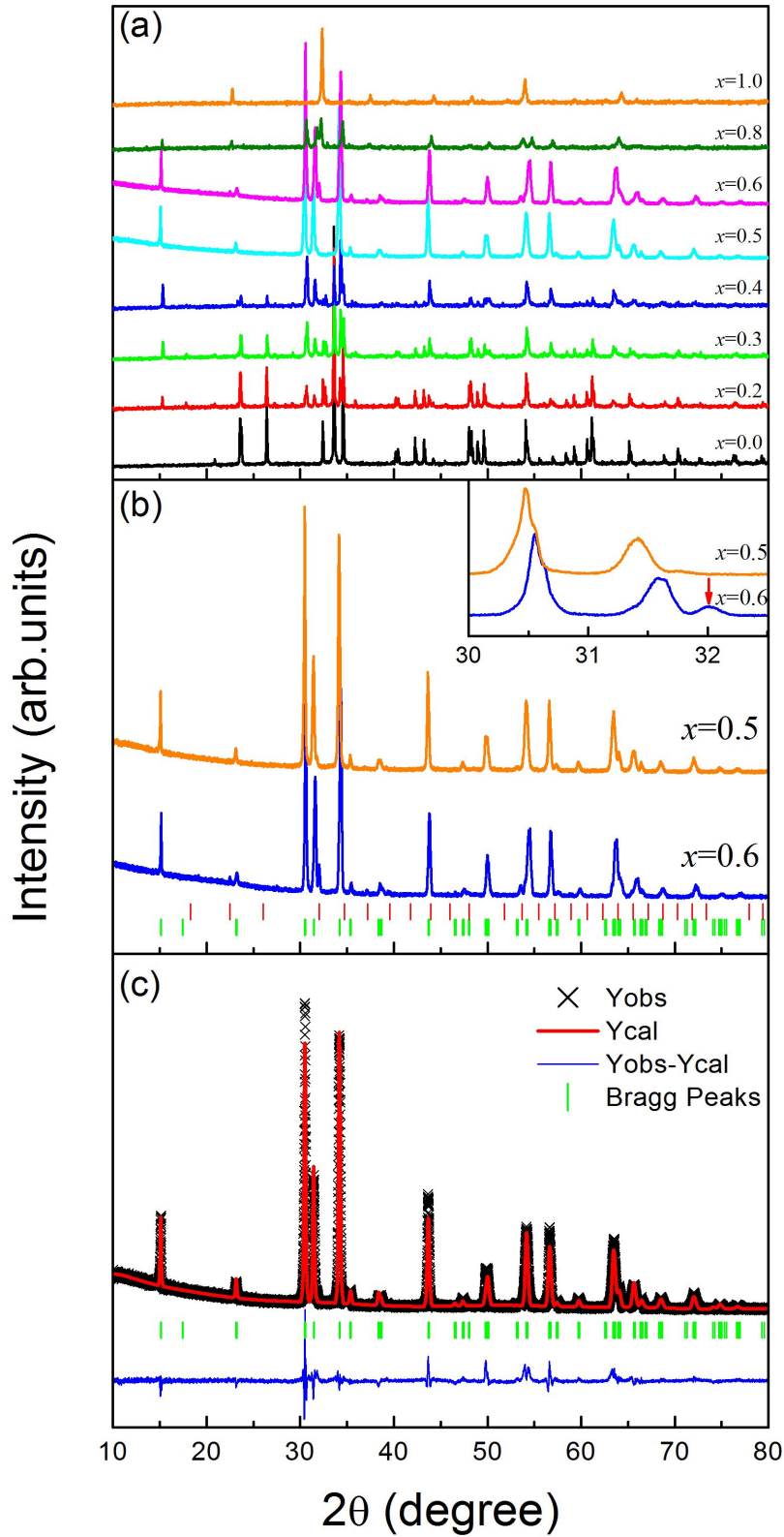

**Fig. 4.** (Color online) (a) X-ray diffraction profiles of $Lu_{1-x}Sc_xFeO_3$ ($0 \leqslant x \leqslant 1$) samples. (b) The slowly-scanned XRD patterns of samples $x=0.5$ and $0.6$. The inset shows the magnified view of peaks, implying slight content of $ScFeO_3$ impurity in sample $x=0.6$. (c) Refinement of XRD pattern of $x=0.5$.

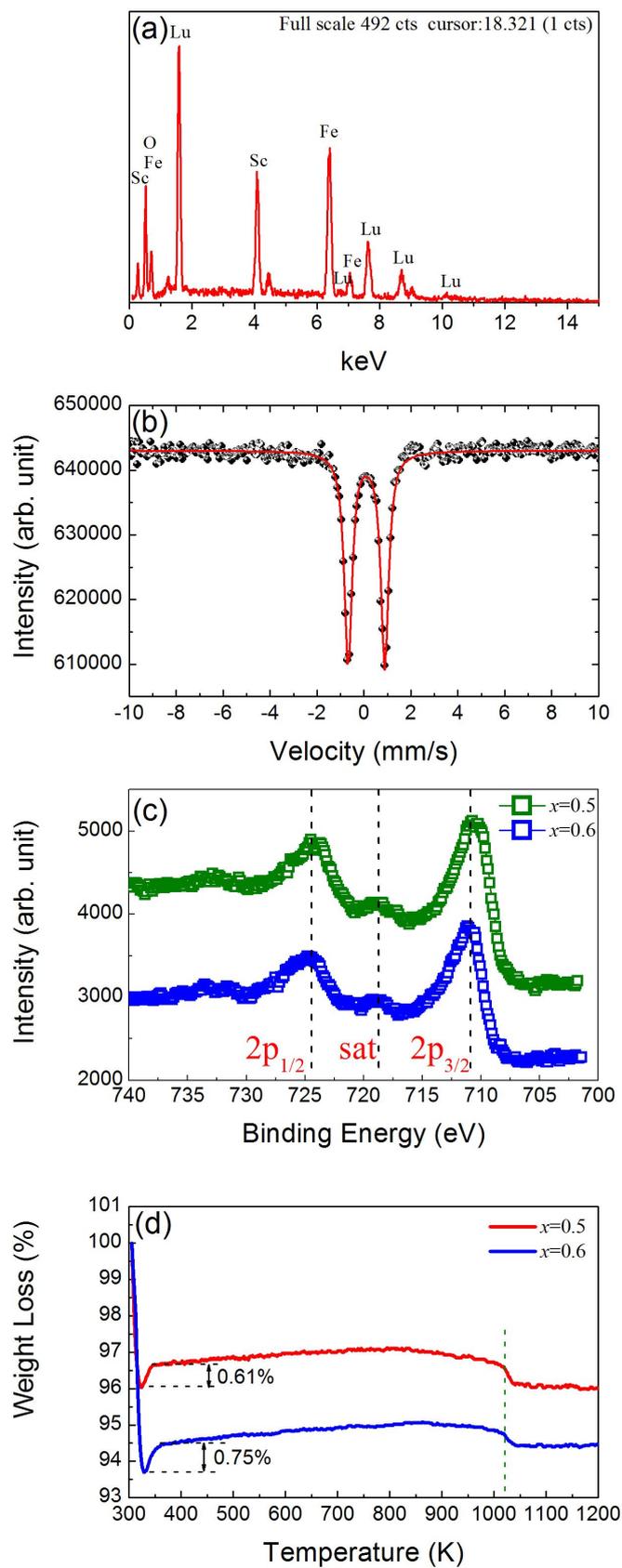

**Fig. 5.** (Color online) The cation/anion non-stoichiometry and Fe valence state analysis of $Lu_{0.5}Sc_{0.5}FeO_3$ by (a) EDS; (b) Mössbauer spectrum; (c) XPS; (d) TG measurements.

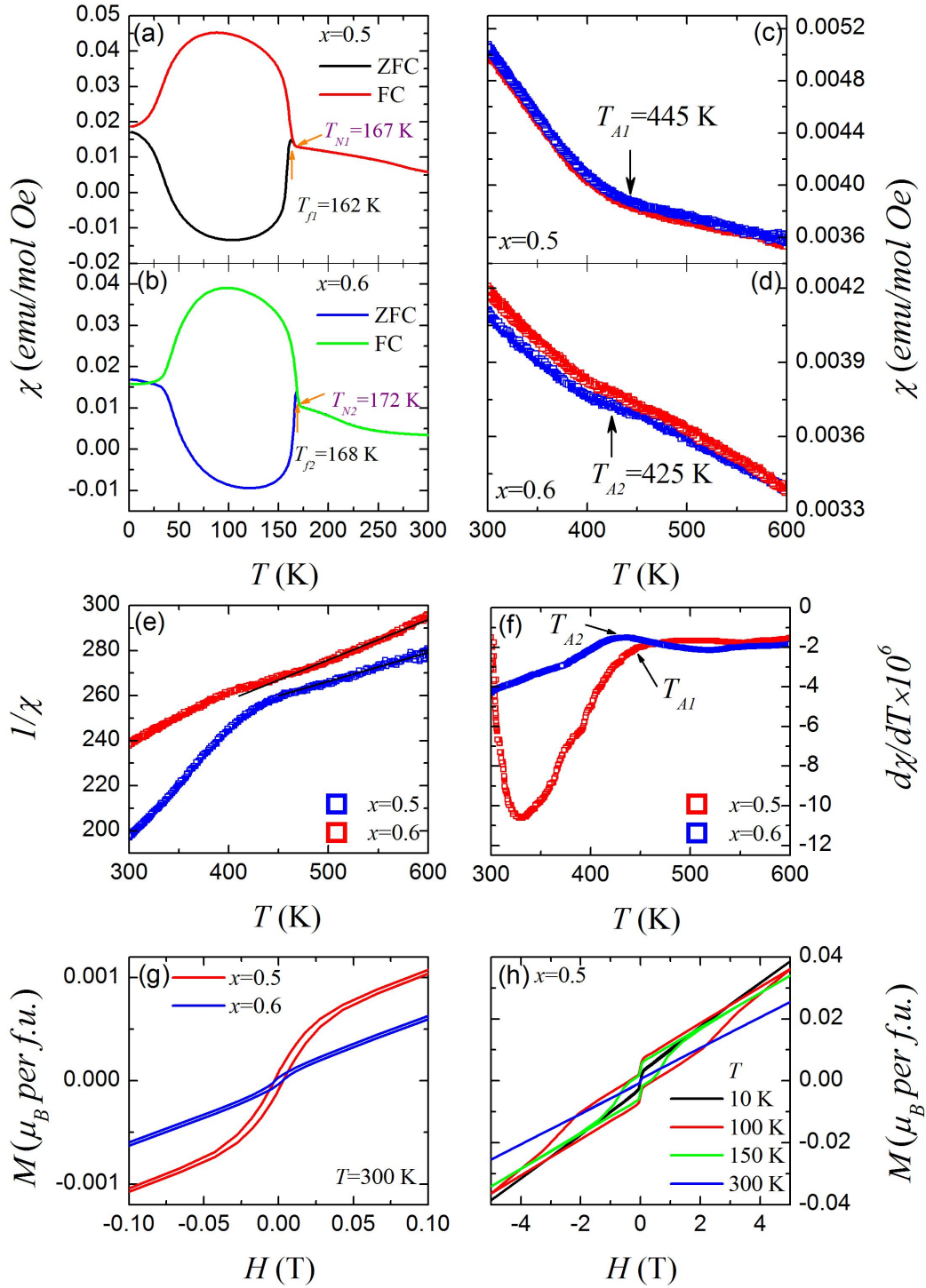

**Fig. 6.** (Color online) Temperature dependence of magnetic susceptibility $\chi$ of sample (a) $x$=0.5, and (b) $x$=0.6 under ZFC and FC modes. The high temperature $\chi(T)$ data of sample (c) $x$=0.5, and (d) $x$=0.6 under ZFC mode. (e) Temperature dependence of the inverse magnetic susceptibility of $x$=0.5 and 0.6. (f) The temperature dependent d$\chi$/d$T$ data of sample $x$=0.5 and $x$=0.6. (g) Magnetic field dependence of magnetization $M$ of samples $x$=0.5 and 0.6 at 300 K. (g) $M$-$H$ loops measured at $T$=10, 100, 150, and 300 K for sample $x$=0.5.

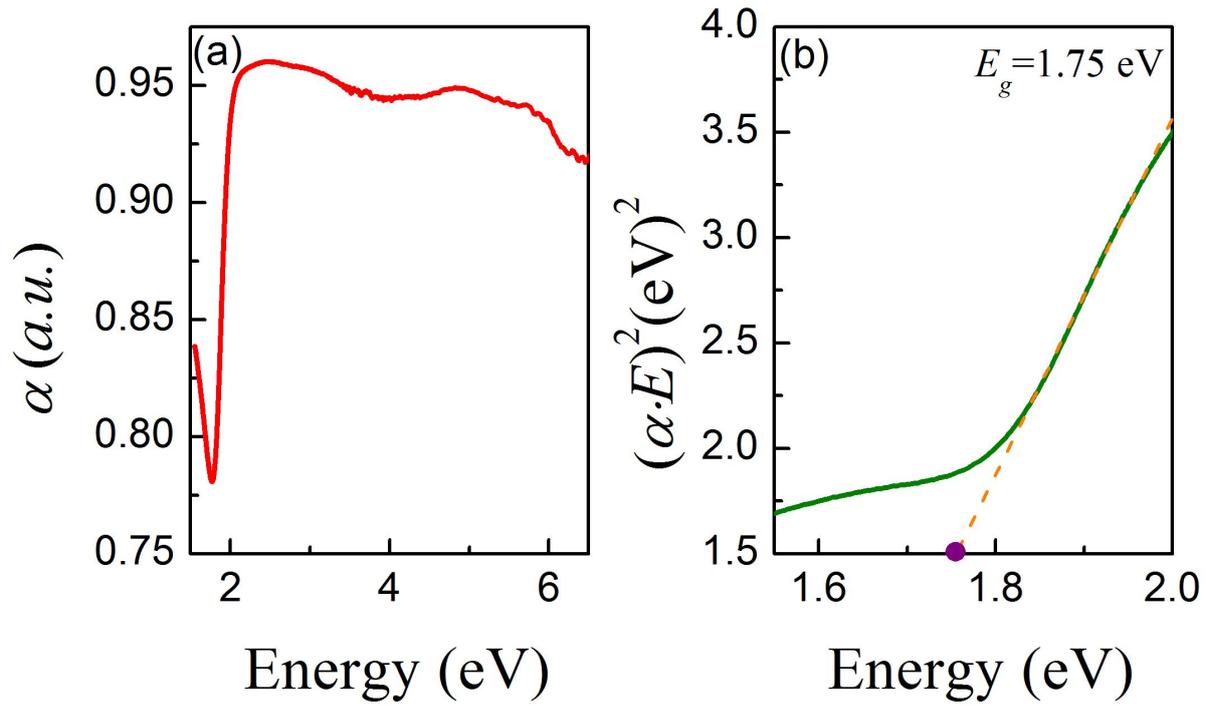

**Fig. 7.** (Color online) (a) Optical absorbance $\alpha$ of $Lu_{0.5}Sc_{0.5}FeO_3$ as a function of phonon energy ($E$). (b) $(\alpha E)^2$ as a function of phonon energy, which indicates an optical bandgap of 1.75 eV.

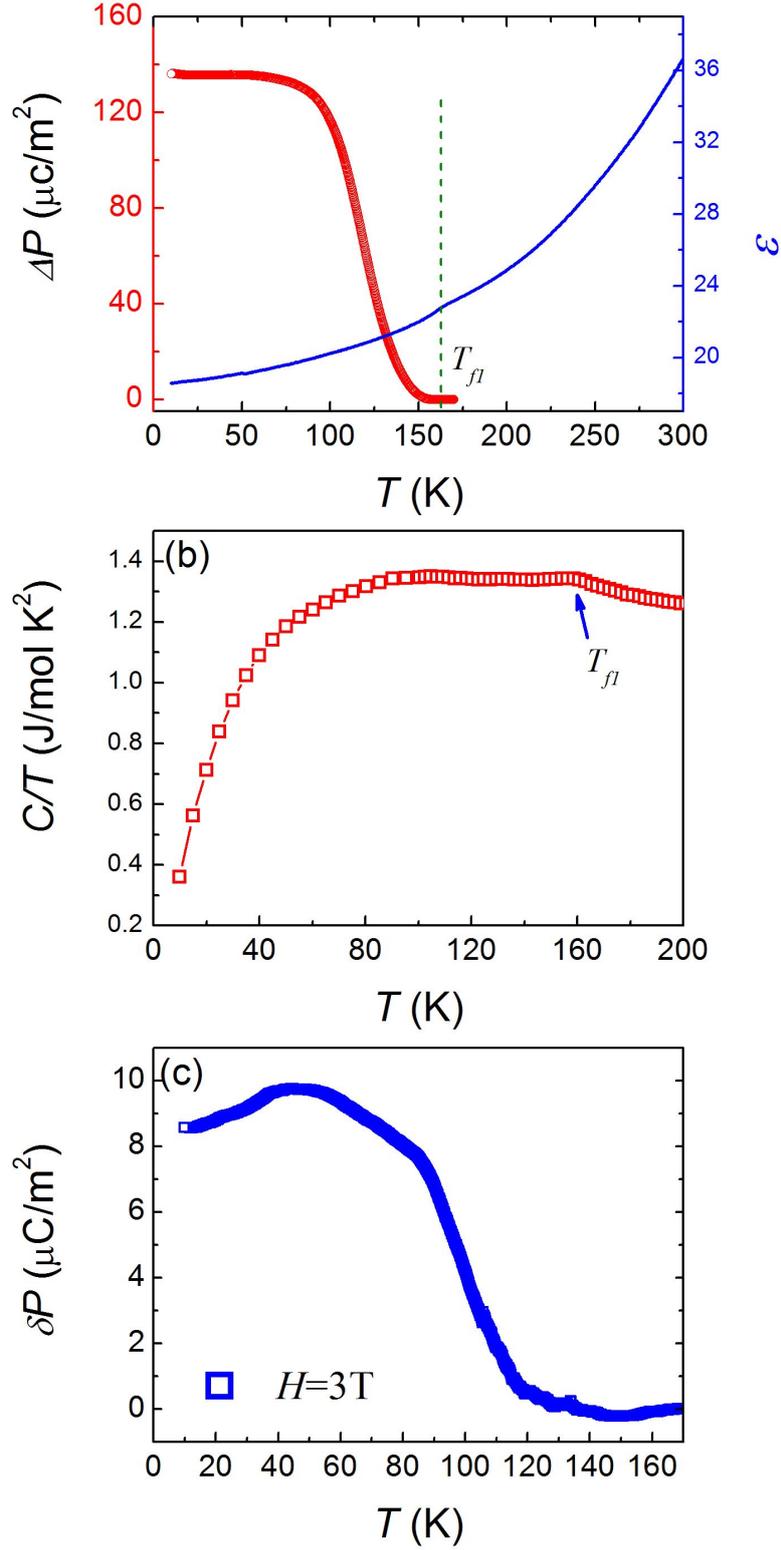

**Fig. 8.** (Color online) (a) Temperature dependence of the evaluated polarization $\Delta P$ and dielectric constant ($\varepsilon$) (b) The specific heat divided by temperature ($C_p/T$) at $H$=0. (c) The variation of polarization $\delta P = P(3\,\text{T}) - P(0\,\text{T})$ as a function of temperature.